\documentclass[12pt,preprint]{aastex}

\def\chan{$\it{Chandra}$}
\def\g292{G292.2$-$0.5}
\def\j1119{PSR~J1119$-$6127}

\received{2007 October 1}
\begin{document}
\title{
Using $Chandra$ to Unveil the High-Energy Properties of the High-Magnetic Field Radio Pulsar J1119$-$6127}

\author{Samar Safi-Harb\altaffilmark{1,2}, Harsha S. Kumar\altaffilmark{1}}

\altaffiltext{1}{Physics and Astronomy Department, University of Manitoba, Winnipeg, MB, R3T 2N2, Canada}

\altaffiltext{2}{Canada Research Chair; samar@physics.umanitoba.ca}

\begin{abstract}
PSR J1119$-$6127 is a high magnetic field (B=4.1$\times$10$^{13}$~Gauss), 
young ($\leq$1,700 year old), and slow (P=408 ms) radio pulsar associated
with the supernova remnant (SNR) G292.2$-$0.5.
In 2003, $Chandra$ allowed the detection of the X-ray counterpart of the radio pulsar,
and provided the  first evidence for a compact and faint pulsar wind nebula (PWN).
We here present new $Chandra$ observations 
which allowed for the first time an imaging and spectroscopic study 
of the pulsar and PWN independently of each other.
The PWN is only evident in the hard band (above $\sim$2 keV) and consists
of jet-like structures extending to at least 7$^{\prime\prime}$ from the pulsar, with
the southern `jet' being longer than the northern `jet'.
The spectrum of the PWN is  described by a power law with a photon index $\Gamma$$\sim$1.1
for the compact PWN and $\sim$1.4 for the southern long jet (at a fixed column
density $N_H$=1.8$\times$10$^{22}$~cm$^{-2}$),
and a total luminosity $L_x$(0.5--7.0 keV)$\sim$4$\times$10$^{32}$~ergs~s$^{-1}$,
at an assumed distance of 8.4 kpc.
The pulsar's spectrum is clearly softer than the PWN's spectrum. 
We rule out a single blackbody model for the pulsar, and present the first
evidence of non-thermal (presumably magnetospheric)
emission that dominates above $\sim$3~keV.
A two-component model consisting of a power law component (with photon index $\Gamma$$\sim$1.5--2.0) 
plus a thermal component provides the best fit. 
The thermal component accounts for the soft emission and 
can be fit by either a blackbody model with a temperature $kT$$\sim$0.21~keV,
or a neutron star atmospheric model with a temperature $kT$$\sim$0.14~keV.
The efficiency of the pulsar in converting its rotational power, $\dot{E}$, 
into non-thermal X-ray emission from the pulsar and PWN is $\approx$5$\times$10$^{-4}$,
comparable to other rotation-powered pulsars with a similar $\dot{E}$. 
We discuss our results in the context of the X-ray manifestation of high-magnetic field radio pulsars
in comparison with rotation-powered pulsars and magnetars.
\end{abstract}


\section{Introduction}
A pulsar wind nebula (PWN) is a bubble of relativistic particles and magnetic field,  
powered by the rotational energy of a young energetic pulsar.
The PWN's non-thermal emission arises from synchrotron radiation of the high-energy particles injected by 
the pulsar and accelerated at the site where the  relativistic pulsar wind
is decelerated by the  confining medium.
As such, PWNs represent an ideal laboratory to study the physics of neutron stars, 
particle acceleration, and relativistic shocks.

The Crab nebula has been viewed for almost three decades as the
canonical example of how a fast rotation-powered pulsar dumps 
its energy into its surroundings.
However, the past several years have witnessed a synergy of 
high-energy  observations pointing to 
a growing class of compact objects which manifest
themselves differently from the Crab. These include 
the anomalous X-ray pulsars (AXPs) and the soft gamma-ray repeaters (SGRs),
commonly believed to be highly magnetized neutron stars, or
magnetars ($B$$\sim$10$^{14}$--10$^{15}$~G), 
powered by magnetic field decay rather than by rotation, 
and characterized by relatively slow periods (5--12 seconds;  see Woods and Thompson 2006 for a review). 
So far, there is no evidence of PWNs surrounding these objects.
Furthermore, they were believed to be radio quiet, however 
recently radio emission has been observed from the transient magnetar
XTE J1810$-$197 (Halpern et al. 2005, Camilo et al. 2006) 
and 2-second radio pulsations have been discovered
from a new magnetar (1E 1547.0$-$5408, Camilo et al. 2007).

Interestingly, there are about half a dozen radio pulsars with spin and magnetic field properties 
intermediate between the `classical' Crab-like pulsars\footnote{A Crab-like pulsar here refers
to a rotation-powered pulsar powering a PWN}
 and the magnetars:
their spin periods range from a fraction of a second to $\sim$7~seconds, and
their magnetic field\footnote{the magnetic field is derived 
assuming magnetic dipole radiation and is related to the spin period, 
$P$, and the spin-down period derivative, $\dot{P}$, as:
$B$=3.2$\times$10$^{19}$$\left(P \dot{P}\right)^{1/2}$ Gauss}
is close to the QED value of  4.4$\times$10$^{13}$ Gauss
(the field at which the electron's rest mass energy equals its 
 cyclotron energy).
The question whether these pulsars should display more magnetar-like or Crab-like properties is 
still under debate (see e.g. the references in Table~3). X-ray observations targeted to
study these objects offer the best window to address 
this question and better understand their emission properties.
The search for and study of any associated PWN should in particular shed light on the nature of their high-energy emission in comparison with the other pulsars.

PSR~J1119$-$6127 is a member of this small, yet growing, class of objects.
It was discovered in the Parkes multi-beam pulsar survey (Camilo et al. 2000).
 It has a rotation period $P$ of 408~ms, a large period derivative $\dot{P}$ of 4$\times$10$^{-12}$, 
a characteristic age $\frac{P}{2\dot{P}}$ of 1600~yrs, 
a surface dipole magnetic field $B$ of 4.1$\times$10$^{13}$~G, 
a spin-down luminosity $\dot{E}$ of 2.3$\times$10$^{36}$ ergs~s$^{-1}$,
and a braking index of 2.9 implying an upper limit on its age of 1,700 years
(under standard assumptions of simple spindown).
No radio PWN was detected around the pulsar despite its youth (Crawford et al. 2001).
The pulsar was found within the supernova remnant (SNR) G292.2$-$0.5
of 15$^{\prime}$ in diameter, discovered with the 
Australia Telescope Compact Array (ATCA) (Crawford et al. 2001), 
and subsequently detected with X-ray observations acquired with the 
$ROSAT$ satellite and the Advanced Satellite for Cosmology and Astrophysics, $ASCA$ (Pivovaroff et al. 2001).

The X-ray counterpart to the radio pulsar was first resolved with
$Chandra$, which also revealed for the first time evidence for 
a compact (3$^{\prime\prime}$$\times$6$^{\prime\prime}$) PWN
(Gonzalez \& Safi-Harb, 2003; hereafter GSH03). The pulsar was subsequently 
detected with \textit{XMM-Newton} and was found to exhibit thermal emission
with an unusually  high pulsed fraction of 74$\pm$14\% in the 0.5--2.0 keV range (Gonzalez et al. 2005).
The \textit{XMM-Newton}-derived temperature was also unusually high, in comparison with other pulsars
of similar age or even younger, like the pulsar in the SNR 3C~58 (Slane et al. 2002).

While a spectrum of the pulsar has been obtained with the previous $Chandra$ and $XMM$-$Newton$ observations, the $Chandra$ spectrum suffered from poor statistics and it wasn't possible to extract a spectrum from the PWN; and the $XMM$-$Newton$ spectrum of the pulsar was contaminated by the PWN's spectrum due to the large point spread function (PSF) of $XMM$-$Newton$.

In this paper, we present new observations of the pulsar obtained with
$Chandra$ and  targeted to constrain the properties of the pulsar and associated PWN.  
In a follow-up paper, we present the study of the associated SNR, 
to complement the previous study of the western side of the remnant by Gonzalez \& Safi-Harb (2005).

The distance to the pulsar was determined to be in the 2.4--8.0 kpc range (Camilo et al. 2000).
Using the fitted X-ray column density to the pulsar and that to its associated SNR,
GSH03 estimated a distance in the 4.0--12.6 kpc range, but the location of the source 
with respect to the Carina spiral arm (Camilo et al. 2000) puts an upper limit on the distance 
of $\sim$8~kpc. The Cordes-Lazio Galactic free electron density model (NE2001, Cordes \& Lazio 2002)
 suggests a much larger distance of 16.7~kpc, while Caswell et al. (2004) estimate a distance of 8.4 kpc using neutral hydrogen absorption measurements to the SNR. In this work, we adopt the distance to the SNR of 8.4~kpc, and scale all derived
quantities in units of $d_{8.4}$=$\frac{D (kpc)}{8.4kpc}$.

The paper is organized as follows. In \S2, we summarize the $Chandra$ observations. 
In \S\S 3 and 4, we present our imaging and spectroscopic results, respectively. 
Finally, in \S5 we present our discussion and conclusions.

\section{Observations and Data Analysis}
PSR~J1119$-$6127 was observed with \chan\  in cycle 5 on 2004 November 2--3 (obsid 6153), and October 31--November 1 (obsid 4676). We combine these new data with the observation conducted during cycle 3 (obsid 2833, GSH03) in order to improve on the statistics. All three observations were positioned at the aimpoint of the back-illuminated S3 chip of the Advanced CCD Imaging Spectrometer (ACIS). X-ray photon events were acquired in the timed-exposure readouts, at a CCD temperature of $-$120$\degr$C and a frame readout time of 3.2 sec,  and telemetered to the ground in the `Very Faint' mode.
The roll angles were chosen such as the AO-3 data covered the western side of the SNR and the
AO-5 data covered the eastern side (see Fig.~1).
The data were then reduced using standard CIAO~4.0 
routines\footnote{http://cxc.harvard.edu/ciao}. 
We have created a cleaned events file by retaining events with
the standard ASCA grades 02346 and rejecting hot
pixels.
As well, the data were filtered for good time intervals,
removing periods with high background rates or unstable aspect.
The resulting effective exposure time was 56.8~ksec for obsid 2833,
60.5~ksec for obsid 4676, and 18.9~ksec for obsid 6153,
yielding a total effective exposure time for the 3 combined observations
covering the field around the pulsar of 136.2~ksec.  

\section{Imaging results}

\subsection{G292.2-0.5}

Figure~1 shows the $Chandra$ cycle 3 (covering the western side of the remnant) and cycle 5 (covering the eastern side of the remnant) observations overlayed on the $ASCA$ hard band (2--10 keV) image, and with the radio extent of the SNR shown as a black circle overlayed. 
The central circle denotes the position of the pulsar.
The image illustrates the power of $Chandra$ in resolving the X-ray emission
and shows the overlap between the $Chandra$ observations leading to a deep 
exposure of the pulsar and central regions of the SNR.
The pulsar and surrounding region will be the focus of this paper. The properties of the diffuse
emission from the SNR 
 will be presented in a separate paper.

Figure~2 shows the combined $Chandra$ tri-color image, zoomed on the pulsar region, and obtained as follows:
 the data were firstly divided into individual images in the soft 
(0.5--1.15 keV, red), medium (1.15--2.3 keV, green), and hard (2.3--10.0 keV, blue) bands
(the energy boundaries were chosen to match those in GSH03).
Individual background images were then obtained from the blank-sky data sets available in CIAO.
For each observation, background images were reprojected to match the corresponding events file. The background images were then divided into the same energy bands mentioned above, and then subtracted from their corresponding
 source+background image. The resulting images were subsequently
adaptively smoothed\footnote{the `csmooth' CIAO tool is based on the
`asmooth' algorithm (Ebeling et al. 2006) which has the advantage of
visually enhancing extended features in noise-dominated regions. We note
that the compact PWN and the southern jet display a similar morphology
using a simple Gaussian smoothing.} using a Gaussian with $\sigma$=$1''$
for significance of detection $>$5 and up to $\sigma$=$5''$ for
significance down to 2. The individual background-subtracted images were
then combined to produce the image shown in Figure~2. 
 As in our previous X-ray study (GSH03), the X-ray emission can be resolved into several X-ray sources 
(tabulated in Table~2 of Gonzalez \& Safi-Harb 2005), surrounded by diffuse emission from the SNR interior.  As well, the pulsar and associated PWN immediately stand out as a hard (blue in color) source.

\subsection{PSR~J1119$-$6127 and its PWN}
 The peak of the X-ray emission from the pulsar is positioned at: $\alpha_{J2000}$=11$^h$19$^m$14$\fs$26, $\delta_{J2000}$ = 
$-$61$\degr$27$^{\prime}$49$\farcs$3, which match within error 
the radio coordinates of the pulsar:
$\alpha$$_{J2000}$ = 11$^{h}$19$^{m}$14$\fs$3 and $\delta$$_{J2000}$ = $-$61$\degr$27$^\prime$49$\farcs$5 (with 0.3$^{\prime\prime}$ error).

In order to compare the source's spatial characteristics with $Chandra$'s PSF, 
we performed a two-dimensional spatial fit to the new $Chandra$
data using the SHERPA software\footnote{http://cxc.harvard.edu/sherpa/}.  
We created images of the source in the soft (0.5-1.15 keV), medium (1.15-2.3 keV),
and hard (2.3-10 keV) bands. We subsequently created normalized PSF images
at the energies characteristic of the
source's energy histogram. These images were then
used as convolution kernels when fitting.
The soft band image yielded a FWHM value of 0.8$^{\prime\prime}$ consistent
with a point source, the
medium band image gave a slightly higher value of 0.9$^{\prime\prime}$,
while the hard band was best
described by an elliptical Gaussian function with FWHM  0.9$^{\prime\prime}$$\times$1.2$^{\prime\prime}$,
confirming the extended nature of the PWN in the hard band,
a result that is consistent with GSH03.

In Figures 3 and 4, we show zoomed in images of the PWN.
Figure~3 shows  the raw soft band (0.5--2 keV) and hard band (2--10 keV) images centered on
the pulsar, and  Figure~4 shows
the same images adaptively smoothed using a Gaussian with $\sigma$=1$^{\prime\prime}$ for
a significance of detection greater than 5 and up to $\sigma$=3$^{\prime\prime}$ for a
significance of detection down to 2.
As mentioned earlier, the PWN structures (further discussed below) are also visible using a simple Gaussian smoothing.
We verified that they are also evident in the hard band images for the individual observations.
Furthermore, using the $celldetect$ CIAO routine, we did not find any point sources within
the PWN, thus ruling out the contamination of the PWN structures by background sources.

The images clearly show that the PWN is harder than the pulsar 
and can be resolved into elongated `jet'-like
structures extending to at least 7$^{\prime\prime}$ north and south of the pulsar.
This confirms the evidence of a hard PWN associated with this
high-magnetic field pulsar (GSH03) but also shows that the PWN is larger than originally thought. 
The PWN is elongated in the north-south direction, with the  northern extension (which we refer to as a  `jet')  appearing tilted to $\sim$30$\degr$ west of north, and the southern jet being more elongated and
$\sim$2$\times$ brighter than the northern `jet'.
In Figure~5, we highlight the first evidence for a more extended ($\gtrsim$20$^{\prime\prime}$)  
faint and hard jet-like structure southwest of the pulsar. 
 We accumulate  100$\pm$23 background-subtracted source counts from this region, which on top of
99$\pm$21 background counts (normalized to the source area), 
imply a $\sim$7$\pm$2$\sigma$ detection (0.5--7.0 keV). 
It is not clear whether this structure is
an extension of the southern compact jet seen closer to the pulsar or 
an SNR filament. 
However, its morphology (in particular its knotty structure which is similar to
Vela's long jet, Kargaltsev et al. 2003) and spectrum (see \S4.2) suggest an association with
\j1119's PWN, a result that can be confirmed with a deeper exposure. 
We note that while the \textit{XMM-Newton} MOS2 image shows a hint of a southern extension,
the large and distorted PSF of the MOS2 camera does not allow a confirmation of
this jet (M. Gonzalez 2008, private communication).

In the following, we refer to the PWN region shown in Figure~4, right (size $\sim$6$^{\prime\prime}$ in the east-west 
direction by $\sim$15$^{\prime\prime}$ in the north-south direction ) 
as the `compact PWN', and to the extended jet-like structure southwest
of the pulsar (size $\sim$6$^{\prime\prime}$$\times$20$^{\prime\prime}$, see Fig.~5) as the `southern jet'.
We note here that it's possible that the compact PWN  corresponds to a torus viewed edge-on.
However, the southern extension (if proven to be a jet associated with the pulsar) would falsify this
interpretation as the jets are expected to be perpendicular to the torus.
We further discuss these features in \S4.2 and \S5.2.

\section{Spatially resolved spectroscopy}
The spectral analysis was performed using the X-ray spectral fitting package, 
XSPEC v12.4.0\footnote{http://xspec.gsfc.nasa.gov}, 
and constrained to the 0.5--7.0 keV band as the spectra provided little useful information outside this range. 
The pulsar's spectrum was grouped with a minimum of 20~counts per bin, while the PWN's spectrum was binned using a minimum of 10 counts~per~bin.
Errors are at the 90\% confidence level throughout the paper. 

\subsection{PSR~J1119$-$6127}
The spectrum of \j1119 was extracted using a circle centered at the position of the pulsar (\S3) 
and with a radius of 2.5$^{\prime\prime}$, thus encompassing 90\% of the encircled energy\footnote{http://cxc.harvard.edu/proposer/POG/html/ACIS.html}.  The background was extracted from a ring centered at the pulsar, and extending  from 3$^{\prime\prime}$ to 4$^{\prime\prime}$. This has the advantage of minimizing contamination by the PWN.
We note however that the PWN contamination is negligible as its surface brightness is nearly two orders
of magnitude smaller than the pulsar's, and so the spectral results are practically
independent of the background subtraction region (see e.g. Safi-Harb 2008 for a preliminary analysis of the pulsar's spectrum
using a source-free background region taken from the S3 chip\footnote{The tabulated values in Safi-Harb (2008)
correspond to binning the data with a minimum of 10 counts per bin, whereas the data here are binned by a minimum of 20 counts per bin.}).
The background-subtracted pulsar count rate  is 
(5.2$\pm$0.3)$\times$10$^{-3}$ counts~s$^{-1}$ for the 2833 observation,
(5.8$\pm$0.3)$\times$10$^{-3}$ counts~s$^{-1}$ for the 4676 observation, and
(3.7$\pm$0.5)$\times$10$^{-3}$ counts~s$^{-1}$ for the 6153 observation,
thus accumulating $\sim$715$\pm$27 total counts from the pulsar.
To model the emission from the pulsar, we used
a blackbody model to account for any surface thermal emission from the neutron star
and a power law to account for any magnetospheric emission. 
A power law model yields a column density $N_H$=(1.2$^{+0.4}_{-0.3}$)$\times$10$^{22}$~cm$^{-2}$, a steep photon index $\Gamma$=3.0$\pm$0.5,  and an unabsorbed flux of
(1.7$^{+1.3}_{-0.7}$)$\times$10$^{-13}$~ergs~cm$^{-2}$~s$^{-1}$,
with a reduced chi-squared value $\chi^2_{\nu}$=1.356 ($\nu$=31 dof).
A blackbody model does not yield an acceptable fit
($\chi^2_{\nu}$=2.166, $\nu$=31 dof); and as
 shown in Figure~6, it does not account for the emission above $\sim$3~keV.
Adding a power law component to the blackbody model improves the fit significantly,
and yields an excellent fit with the following parameters: 
$N_H$=(1.8$^{+1.5}_{-0.6}$)$\times$10$^{22}$~cm$^{-2},
 \Gamma$=1.9$^{+1.1}_{-0.9}$,  $kT$=0.21$\pm$0.01~keV, $\chi^2_{\nu}$=0.948 ($\nu$=29 dof),
with the luminosity of the power law component representing $\sim$30\% of the luminosity
of the blackbody component (or $\sim$23\% of the total pulsar's luminosity)
in the 0.5--7.0 keV range.
Figure~7 shows the blackbody+power law model fit, and Table~1 summarizes the best fit parameters. 

The thermal  and soft component is  equally well fit with a neutron star magnetized (B=10$^{13}$~G) 
atmospheric model (the NSA code in XSPEC, Zavlin et al. 1996). The parameters for the blackbody and atmospheric model
are however different, and as expected the NSA model yields a lower temperature than the 
blackbody model. Both models require a hard photon index ($\Gamma$$\sim$1.5--2.0)
for the power law component.
As shown in Table~1,  the
non-thermal unabsorbed flux amounts to $\sim$30--35\% of the thermal flux.
In the NSA model, the distance was fixed to 8.4~kpc. 
We attempted to fit for this distance, however we 
find that it is poorly constrained (ranging from 2.6 to 30 kpc).
The best fit parameters are: 
$N_H$=1.6$\times$10$^{22}$~cm$^{-2}$,
$\Gamma_{PL}$=1.5, $T_{NSA}$=1.6$\times$10$^6$~K, and $D$=8.4~kpc.
Therefore we restrict our discussion below to the fit with the fixed and reasonable distance of 8.4~kpc.

\subsection{The PWN}
The compact PWN's spectrum was extracted from the elongated nebula ($\sim$6$^{\prime\prime}$$\times$15$^{\prime\prime}$ )(see Fig.~4, right) with the pulsar region removed.
The background was extracted from a nearby source-free region. 
The total number of background-subtracted counts accumulated from the PWN is only $\sim$103$\pm$17~counts in the 0.5--7.0 keV range,
therefore restricting our ability to constrain the spectral parameters or to perform a spatially resolved spectroscopy of the northern and southern extensions. However, fixing the column density
to 1.8$\times$10$^{22}$~cm$^{-2}$ (the best fit value for the pulsar in the BB+PL fit, Table~1),
we find that a power law model yields a hard photon index $\Gamma$=1.1$^{+0.9}_{-0.7}$, 
$\chi^2_{\nu}$=1.23 ($\nu$=17 dof), and a luminosity
$L_x$(0.5--7.0 keV)$\sim$(1.6$^{+2.3}_{-0.9}$)$\times$10$^{32}$~$d_{8.4}^2$  ergs~s$^{-1}$.
Allowing $N_H$ to vary between (1.2--3.3)$\times$10$^{22}$~cm$^{-2}$ (the range allowed by the PL+BB model fit to the pulsar), the PWN's index is $\Gamma$=0.6--2.9, typical of PWNs.

We subsequently extracted a spectrum from the southern `jet' (region shown in Fig.~5)
and used the same background region as for the compact PWN.  
As mentioned earlier, we accumulated a total of $\sim$100$\pm$23 source counts 
from this region in the 0.5--7.0 keV range, again too small to constrain the parameters or
perform a spatially resolved spectroscopy along this feature. Fixing the column density
to  1.8$\times$10$^{22}$~cm$^{-2}$, the power law model yields a photon index of $\Gamma$=1.4$^{+0.8}_{-0.9}$,  $\chi^2_{\nu}$=1.16 ($\nu$=17 dof), and a luminosity of 
$L_x$(0.5--7.0 keV)$\sim$(2.1$^{+3.3}_{-1.2}$)$\times$10$^{33}$~$d_{8.4}^2$ ergs~s$^{-1}$. 
These values are not unusual for PWN structures.
A thermal bremsstrahlung model is rejected based on the unrealistically  high temperature ($kT$$\geq$10~keV).
Experimenting with
two-component models was not possible given the insufficient number of counts.
In Table~2, we summarize the best fit power law model results. 

\section{Discussion}

\subsection{The pulsar}

For the first time, we have resolved the pulsar's spectrum and determined its properties independently of the PWN. We note here that the hard X-ray component in the \textit{XMM-Newton} spectrum has been attributed entirely to the emission from the PWN (Gonzalez et al. 2005). We believe that this component is contaminated by the hard X-ray emission from the pulsar as \textit{XMM}'s PSF does not allow resolving both components. In fact, an on-off pulse spectroscopic analysis of the \textit{XMM}'s spectrum  shows evidence of pulsed power law emission at the 2$\sigma$ level  (M. Gonzalez 2008, private communication).

The single blackbody model is rejected  (see \S4.1 and Fig.~6). As shown in Table~1,
the best fit model requires two components: a thermal component
described by a BB or NSA model, and a power law component with a photon index $\Gamma$$\sim$1.5--2.0.
The column density is model dependent. 
GSH03 derived an $N_H$=(9$^{+5}_{-3}$)$\times$10$^{21}$~cm$^{-2}$ for the pulsar. 
Using \textit{XMM}, Gonzalez et al. (2005) derived a higher  $N_H$ in the (1.3--2.7)$\times$10$^{22}$~cm$^{-2}$ range 
(depending on the model used). Our best fit 
 two-component models shown in Table~1, which adequately account for the hard (above 3 keV) emission,
yield a column density of (1.4--1.8)$\times$10$^{22}$~cm$^{-2}$ in the NSA+PL model, and
(1.2--3.3)$\times$10$^{22}$~cm$^{-2}$ in the PL+BB model. These ranges bracket the value 
derived using HEASARC's $N_H$ tool\footnote{this tool calculates the
weighted average measurement of the
total Galactic column density towards the source's coordinates based on the Dickey \& Lockman (1990) \ion{H}{1} map.} and overlap
with the values derived previously with $Chandra$ and \textit{XMM}.

Using the thermal model in the two-component fits tabulated in Table~1, we now estimate the
thermal properties of the pulsar. In the blackbody fit,  the radius  is inferred from equating the
isotropic luminosity (4$\pi$$D^2$$F$, where $D$ is the distance to the pulsar and $F$ is the
unabsorbed flux, tabulated in Table~2) with the surface emission from the neutron star 
(4$\pi$$R^2$$\sigma$$T_{eff}$$^4$, where $\sigma$ is the Stefan Boltzmann constant), yielding
$R$=2.7$\pm$0.7~km,
suggesting polar cap emission with a temperature of 
$T_{eff}$=0.21$\pm$0.01~keV=(2.46$\pm$0.12)$\times$10$^6$~K.
The inferred radius is however larger than the conventional polar cap radius: $R_{pc}$(km)$\simeq$0.5$\left(\frac{P}{0.1s}\right)^{-1/2}$=0.25 km (for $P$=408 ms). 
The pulsar's thermal luminosity (assuming isotropic emission) is  $L_x$(0.5--7.0
keV)=(1.87$^{+0.8}_{-0.4}$)$\times$10$^{33}$$d_{8.4}^2$ ergs~s$^{-1}$.
 
 In the atmospheric (NSA) model, the temperature is $T_{eff}$=(1.59$^{+0.35}_{-0.23}$)$\times$10$^6$~K, implying a temperature as seen by an observer of $T_{eff}^{\infty}$=$g_r$$T_{eff}$=(1.22$^{+0.27}_{-0.18}$)$\times$10$^6$~K;
where $g_r$ is the redshift parameter given by $g_r$=(1-2.952$\times$$\frac{M}{R}$)$^{0.5}$=0.766 for a neutron star of mass $M$=1.4 solar masses and radius $R$=10 km. The pulsar's  bolometric luminosity as seen by an observer and assuming thermal emission from the entire neutron star's surface is  $L_{bol}$$^{\infty}$(0.5--7.0 keV) = $g_r^2$$\times$(4$\pi$$R^2$$\sigma$$T_{eff}^4$)=(2.67$\pm$1.9)$\times$10$^{33}$$d_{8.4}^2$ ergs~s$^{-1}$. 
These estimates differ from the values inferred by Gonzalez et al. (2005), 
mainly because the distance was fixed here at 8.4~kpc, while it was fit for with \textit{XMM}.
Our values for both the blackbody and NSA model are however consistent with those derived independently by Zavlin (2007a).

 The thermal emission from the pulsar was already discussed by Gonzalez et al. (2005) and 
 Zavlin (2007a) who point out that the large pulsed fraction detected in the soft band is hard to
 interpret in either the blackbody or the NSA model (assuming uniform surface emission).
The large pulsed fraction indicates that the thermal emission
is intrinsically anisotropic and that the effective temperature derived in the NSA model
should be used as the mean surface temperature (Zavlin et al. 2007b).
We here point out to a puzzle related to the interpretation of the hard non-thermal pulsar emission, seen for the
 first time with the $Chandra$ observations presented here. The non-thermal emission
 is presumably magnetospheric emission as seen in other young active pulsars.
 Such an emission is expected to be pulsed. As mentioned earlier, \textit{XMM} could
not resolve the pulsar's emission from the PWN, however an on-off pulse
spectroscopic study of the \textit{XMM} spectrum shows a hint of 
pulsed power law emission from the pulsar.
A long observation with $Chandra$ in the timing mode and/or a deeper \textit{XMM-Newton} observation will confirm and constrain the pulsed component in the hard band.

\subsection{The PWN}

The PWN is well described by a power law model with a hard photon index 
$\sim$1.1 for the compact PWN, and $\sim$1.4 for the southern extended `jet' (for a column density 
fixed at 1.8$\times$10$^{22}$ cm$^{-2}$),
and with a total X-ray luminosity of
$\approx$4$\times$10$^{32}$$D_{8.4}^2$ ergs~s$^{-1}$ (0.5--7 keV). This
represents $\lesssim$20\%  of the pulsar's luminosity, and
$\lesssim$0.02\% of $\dot{E}$, implying an efficiency which is
comparable to PSR B1800$-$21 and other Vela-like pulsars (Kargaltsev et al. 2007). 

PWNs have been observed to have axisymmetric morphologies, including a toroidal structure
and jets along the pulsar's spin axis. The emission from the torus is commonly attributed to
synchrotron emission from the relativistic pulsar wind in the presence of a toroidal magnetic
field in the equatorial plane and downstream of the shock.
Jet-like structures on the other hand are attributed to magnetically collimated winds along
the pulsar's rotation axis.
The collimation and acceleration of the jets is strongly dependent on the pulsar's wind magnetization $\sigma$, defined
as the ratio of electromagnetic to particles flux.
At higher magnetization, equipartition is reached in the close vicinity of the termination shock, and most of the plasma is diverted and collimated into a jet along the polar axis.
Therefore the morphology of the PWN sheds light on the geometry of the pulsar and the
intrinsic parameters of the pulsar wind (see e.g. Gaensler \& Slane 2006 and Bucciantini 2008 for reviews). 

The jet-like structures seen north and south of PSR~J1119$-$6127 could be interpreted as collimated outflows along the pulsar's polar axis, in which case a torus could be expected in the east-west direction. A torus-like structure is however barely resolved and the PWN emission is mostly evident in the form of elongated jets (see Figs.~3--5), suggesting a high-$\sigma$ wind.
In the following, we use the spectral properties of the PWN to infer its intrinsic properties
such as the termination shock, magnetic field, and $\sigma$.

The termination shock site ($r_s$) is estimated by equating the thrust of the pulsar, $\frac{\dot{E}}{4\pi r^2_s c \Omega}$, with the pressure in the nebula, $P$$\approx$$\frac{B_n^2}{4\pi}$; where $\Omega$ is the pulsar's wind solid angle which  accounts for a non-isotropic wind, and $B_n$ is the nebular magnetic field which can be approximated from its equipartition value $B_{eq}$.
To estimate $B_{eq}$, we use the non-thermal properties of the compact PWN (see Table~3), and 
assume the emission region to be a cylinder with radius $r_n$$\leq$3$^{\prime\prime}$=0.12 $d_{8.4}$ pc  and  length of $l$$\sim$15$^{\prime\prime}$=0.61 $d_{8.4}$~pc, resulting in an emitting volume of 
$V$$\sim$8.4$\times$10$^{53}$~$f$$d_{8.4}^3$~cm$^3$; where $f$ is the volume filling factor.
Using a luminosity of $L_x$(0.5--7.0 keV)$\sim$2$\times$10$^{32}$~$d_{8.4}^2$~ergs~s$^{-1}$
and assuming a ratio of 100 for the baryon to electron energy density, we estimate a magnetic
field of $B_{eq}$$\geq$4.5$\times$10$^{-5}$$(f d_{8.4})^{-2/7}$~Gauss. 
The corresponding nebular pressure is  1.6$\times$10$^{-10}$~$(f d_{8.4})^{-4/7}$~ergs~cm$^{-3}$ 
and the termination shock is at a radius $r_s$$\sim$0.06~$(f d_{8.4})^{2/7}$$\Omega^{-1/2}$~pc
(or 1.5$^{\prime\prime}$ at 8.4 kpc), comparable to B1800-21 other PWNs (Kargaltsev et al. 2007). 

We can further check the above estimate by using the thermal pressure from the interior  of the remnant ($\sim$2$n_{e}kT$) as the pressure confining the PWN. Using the  $kT$ and $n_{e}t$ values derived from our spectral fits to the SNR interior (to be presented in our follow up SNR paper),  we derive an SNR internal thermal pressure of $\sim$2.9$\times$10$^{-9}$ ergs~cm$^{-3}$. The resulting shock radius is $r_{s}$$\sim$0.015~pc,  smaller than our estimate above, but consistent for 
$(f d_{8.4})^{2/7}$$\Omega^{-1/2}$=0.25.
At a distance of 8.4~kpc, a termination shock radius of 0.015--0.06 pc translates to a radius of
$r_s$=0$\farcs$4--1$\farcs$5~$d_{8.4}$, which explains
the lack of resolvable wisp-like structures as seen in other (nearby) PWNs.

Using the basic Kennel \& Coroniti (1984) model developed for the Crab nebula, we can further infer 
a rough value for the magnetization parameter $\sigma$$\sim$$\left(\frac{r_s}{r_n}\right)^2$$\sim$0.02--0.25.  This value is larger than that inferred for Crab-like PWNs (see e.g. Petre et al. 2007),
but  supports the picture for strong collimation of jets. We caution however that this estimate
should not be taken at face value and should be only considered as evidence for
a highly magnetized wind, as it was  inferred using the simple model of Kennel \& Coroniti (1984) which 
assumes an isotropic energy flux  in the wind. More recent calculations however show
that if hoop stresses are at work in the mildly relativistic flow, jet collimation can occur in the post shock region and estimates of $\sigma$ using the Kennel and Coroniti model can be overestimated (see e.g. Bucciantini 2008 and references therein for a review).

The corresponding synchrotron lifetime of a  photon with an energy of $E$(keV) is roughly
$\tau_{syn}$$\approx$36$B_{-4}^{3/2}$$E^{1/2}$~$d^{-3/7}_{8.4}$ yr, 
where $B_{-4}$ is the nebular magnetic field in units of 10$^{-4}$ Gauss;
 yielding an average velocity for the high-energy electrons to
 reach the edge of the southern jet of
$\sim$0.093$c$~$B_{-4}^{-3/2}$$E^{-1/2}$~$d^{3/7}_{8.4}$.
This velocity is not unusual and is smaller than the typical shocked wind velocity
of $\sim$0.3$c$.

We interpreted the elongated structures to the north and south of the pulsar as jets that could be aligned with the rotation axis of the pulsar.  The northern extension is shorter in length and is approximately twice as faint as the southern jet. Assuming that these elongated features represent an approaching (south) and receding (north) jet with similar intrinsic surface brightness and outflow velocities, we can use relativistic Doppler boosting to account for the observed difference in brightness (Mirabel \& Rodriguez 1999). For the observed brightness ratio of $\sim$2 and an energy spectral index $-\alpha$$\sim$0.1--0.4 (Table~2), assuming a continuous jet, we estimate a velocity $v\cos\theta$$\sim$0.14--0.16$c$, where $\theta$ is the inclination angle of the flow axis with respect
to our line of sight. 
Since $\cos\theta$$\leq$1 for all inclination angles, this represents a lower limit for the intrinsic  velocity of these features, a result that is consistent with our estimate of the average velocity above.

Elongated jets have been observed in other PWNs, e.g. PSR B1509$-$58 powers a 4$^{\prime}$-long jet (6~pc at a distance of 5.2 kpc),  and Vela powers a 100$^{\prime\prime}$-long  jet (0.14~pc at a distance of 300 pc). Furthermore, velocities of $>$0.2$c$ have been also inferred from X-ray studies of these jets: $v_j$$>$0.2$c$ for B1509$-$58 (Gaensler et al. 2002)  and $v_j$$\sim$(0.3--0.6)$c$ for Vela (Kargaltsev et al. 2003).

\subsection{Comparison of \j1119 to other rotation-powered pulsars and magnetars}

There are currently 7 pulsars including PSR~J1119$-$6127 with magnetar-strength fields 
(B$\gtrsim$B$_c$=4.4$\times$10$^{13}$ G, see Table~3), but which have not been identified as magnetars, except for PSR~J1846$-$0258 which most recently revealed itself as a magnetar (Kumar \& Safi-Harb 2008, Gavriil et al. 2008).  One would ask then whether all other pulsars belonging to this class should be related to magnetars (e.g. by being quiescent magnetars), or whether they represent a distinct population of rotation-powered pulsars.  To address this open issue, we compare the known properties of \j1119 to magnetars and to the `classical' rotation-powered pulsars:

\begin{itemize}
\item{
Unlike magnetars (even in quiescence) and like the classical rotation-powered pulsars, PSR~J1119$-$6127 and its PWN are powered by rotation ($L_x$ is much smaller than its $\dot{E}$)  and the pulsar powers a hard PWN. As well its infrared properties resemble more the rotation-powered pulsars than the magnetars
(Mignani et al. 2007) --  the latter are known to be more efficient IR
emitters.} 
\item{
Like magnetars and some rotation-powered pulsars (see more below), J1119$-$6127 is described by a two-component  (blackbody+power law) model.
PSR~J1119$-$6127's thermal temperature ($kT$$\sim$0.21 keV in the BB+PL model and 0.14 keV in the NSA+PL model) is intermediate between 
that of magnetars ($kT_{BB}$$\sim$0.41--0.67 keV, Woods \& Thompson 2006)
and that of the Vela-like, rotation-powered, pulsars with a thermal component dominating at energies $E$$\leq$2~keV (see Fig.~5 of Zavlin 2007b).
The relatively high temperature of PSR~J1119$-$6127 (in comparison with the rotation-powered pulsars) can be accounted for by a neutron star cooling model with proton
superfluidity in its core and a neutron star mass of $\sim$1.5 M$_{\sun}$ (see Fig.~5 of Zavlin 2007b).
Its photon index (1.0--2.9 in the BB+PL model and 0.7--2.3 in the NSA+PL model, see Table~1) appears relatively hard in comparison to that of most
AXPs in quiescence (see Table~14.1 of Woods \& Thompson 2006) and consistent with that expected from
the rotation-powered pulsars; but also similar to that of three
spectrally-hard magnetars in their quiescent state (SGR~1806$-$20,
SGR~1900+14, and AXP 1E~1841$-$045 having PL indices of 2.0, 1.0--2.5, and 2.0, respectively; Woods \& Thompson 2006).}
%
\item{\j1119's X-ray properties are most similar to two young `Vela-like' pulsars with a similar $\dot{E}$:   PSR~B1800$-$21 and PSR~J1357$-$6429's, rotation-powered pulsars  with a more `typical'  magnetic field (see Table~3 and references therein). Most notably, both pulsars power compact and hard PWNe and  display an $\frac{L_x}{\dot{E}}$ value comparable to \j1119's.
Furthermore, like \j1119, the spectrum of PSR~J1357$-$6429 has a thermal component
that dominates the total flux at energies below 2~keV ($\sim$72\% for PSR~J1357$-$642 versus $\sim$87\% for \j1119)
and is strongly pulsed
despite its lower B-field.
Such a high pulsed fraction ($\geq$50\% at E$\leq$2~keV, Zavlin 2007b) could be explained by intrinsic anisotropy of the thermal
emission formed in a magnetized neutron star atmosphere coupled with a
strong non-uniformity of the surface temperature and magnetic field
 distributions (V. E. Zavlin 2008, private communication; Zavlin 2007b). 
Such an interpretation can also be applied to \j1119 whose thermal spectrum 
is also well fitted with a neutron star atmospheric model  
(see Table~1 and \S5.1; V. E. Zavlin 2008, private communication, Zavlin 2007a).}
\item{
Among the high-B pulsars, PSR~J1119$-$6127 has spin properties (i.e.
$P$, $B$) most similar to those of PSR~J1846$-$0258 (the Kes~75 pulsar).
Both pulsars are also now believed to be at about the same distance of
$\sim$6~kpc.  However PSR~J1846$-$0258 is a non-thermal X-ray pulsar
characterized by a hard power-law index\footnote{we
note here that the column density towards PSR~J1846$-$0258 is at least
twice as high as that towards PSR~J1119$-$6127, thus possibly 
hindering the significant detection of any soft blackbody component
(Kumar \& Safi-Harb 2008).}  ($\Gamma$$\leq$2) and is much more efficient at powering a
bright PWN  ($\frac{L_x}{\dot{E}}$(PWN)$\sim$2\%, see Kumar \& Safi-Harb 2008),  so their X-ray properties are very different.  Furthermore, PSR~J1846$-$0258 has recently revealed itself as a magnetar as  magnetar-like X-ray bursts were detected with $RXTE$  and
its spectrum softened as its flux brightened by a factor of $\sim$6
(Gavriil et al. 2008, Kumar \& Safi-Harb 2008). 
PSR~J1119$-$6127 appears to have a stable spectrum between 2002 (obsid 2833) and 2004 (obsid 4676) with a count rate of $\sim$5.5$\times$10$^{-3}$~counts~s$^{-1}$ (\S4.1).
Obsid 6153 yields a smaller count rate for the pulsar ($\sim$70\% the average count rate) suggesting variability; however if real, we believe that it is not significant within error and given that  1) obsid 6153 was relatively short and taken the day following obsid~4676, and 2) magnetar outbursts normally cause a much brighter enhancement in the X-ray flux. 
It is possible however that \j1119 will one day reveal itself as a magnetar after an occasional burst
driven by its high B-field, just like PSR~J1846$-$0258 recently did.
Monitoring \j1119 and the other high-B pulsars in the radio and X-ray
bands are needed to address this possibility.
 }
\end{itemize}

Given the close resemblance of \j1119 to the other rotation-powered
pulsars with `typical' B-fields and a similar $\dot{E}$, 
we conclude that the global X-ray properties of PSR~J1119$-$6127 are not (at least not entirely) determined by the high B value\footnote{One should also  keep in mind that the actual magnetic field strength can be a factor of few off from the spin-down value, and so pulsars with comparable B may have substantially different actual fields (see Camilo 2008 and references therein.)};  and that $\dot{E}$ in particular, and likely 
the environment which should confine the PWN and therefore would affect its morphology and brightness, play a significant role 
in determining its properties.
Whether \j1119 is a quiescent magnetar that's currently mainly powered by its $\dot{E}$, but
that will one day reveal a magnetar-like identity,  remains to be seen.
 
Finally, our study further suggests that \j1119 is characterized by a highly magnetized wind, possibly explaining the evidence for a small torus and prominent jets.  The deep search for X-ray emission and PWNs around the other high-B pulsars is warranted to address the question whether these pulsars are similarly characterized by highly magnetized winds. Given the small $\dot{E}$ of most known high-B pulsars (Table~3), very long exposures will be needed to be sensitive to PWNs of luminosity $\approx$10$^{-4}$--10$^{-3}$$\dot{E}$.

\acknowledgments
S. Safi-Harb acknowledges support by the Natural Sciences and Engineering Research Council of Canada (NSERC) and the Canada
Research Chairs program. This research  made use of  NASA's Astrophysics Data System, the High-Energy Astrophysics Science
Archive Center operated by NASA's Goddard Space Flight Center, and the ATNF pulsar
database\footnote{http://www.atnf.csiro.au/research/pulsar/psrcat/}. We thank the anonymous referee and F. Camilo for a detailed reading of the manuscript and for useful comments.

\clearpage

\begin{table*}[ht]
\begin{tabular}{l l l l}
\hline
 Parameter & PL  & BB+PL & NSA+PL\tablenotemark{a} \\
\hline
 $ N_{\rm H}~(10^{22}$ cm$^{-2}$) & 1.2$_{-0.3}^{+0.4}$  & 
1.8$_{-0.6}^{+1.5}$ & 1.6$\pm$0.2 \\

$\Gamma$ & 3.0$_{-0.5}^{+0.5}$ & 1.9$_{-0.9}^{+1.1}$ & 1.5$\pm$0.8\\

Norm.\tablenotemark{b} (PL)& 5.7$_{-2.3}^{+4.2}\times10^{-5}$  & 
1.5$_{-0.9}^{+2.3}\times10^{-5}$ & 8.4$_{-5.5}^{+1.5}\times10^{-6}$  \\

$kT_{\rm eff}$(keV) & \nodata & 0.21$\pm$0.01 & 0.14$_{-0.02}^{+0.03}$   \\

 Norm.\tablenotemark{b} (BB)& \nodata  & 3.5$_{-0.8}^{+1.5}\times10^{-6}$ & 
1.42$\times10^{-8}$ (frozen)\\

$F_{\rm unabs}(PL)$\tablenotemark{c} & 1.7$_{-0.7}^{+1.3}\times10^{-13}$ & 
6.7$_{-2.2}^{+0.7}\times10^{-14}$ & 6.1$_{-3.5}^{+3.2}\times10^{-14}$ \\

$F_{\rm unabs}$\tablenotemark{c} (thermal) &\nodata & 2.2$_{-0.5}^{+0.9}\times10^{-13}$ & 
1.8$_{-1.5}^{+1.9}\times10^{-13}$ \\

$L_{\rm X}$\tablenotemark{d} 10$^{33}$ergs~s$^{-1}$  & 1.4$^{+1.1}_{-0.6}$ & 
2.4$^{+0.8}_{-0.5}$& 2.1$^{+0.8}_{-0.6}$ \\

$\chi_{\rm \nu}^2$~(dof) & 1.356 (31) & 0.948 (29)& 0.977 (30) \\

\hline

\end{tabular}
\caption{Spectral fits to the pulsar. All models are modified by interstellar absorption
with a column density $N_{H}$.  PL denotes a power-law model, BB denotes a blackbody, 
and NSA refers to the neutron star H atmospheric model in XSPEC version 12.4.0.}
\tablenotetext{a}{The NSA model assumes a magnetic field
of 10$^{13}$ Gauss, a neutron star radius of 10 km and mass
of 1.4 $M_{\sun}$. The distance is also fixed at 8.4~kpc (thus the normalization of 1.4$\times$10$^{-8}$). }
\tablenotetext{b}{PL normalization at 1 keV in units of photons~cm$^{-2}$~s$^{-1}$~keV$^{-1}$.
BB normalization in units of $L_{39}$/$D_{10}^2$ where $L_{39}$ is the
luminosity in units of 10$^{39}$~ergs~s$^{-1}$ and $D_{10}$ is the distance in units of 10 kpc.}
\tablenotetext{c}{0.5--7.0 keV unabsorbed flux (ergs~cm$^{-2}$~s$^{-1}$).}
\tablenotetext{d}{0.5-7.0 keV total luminosity (assuming isotropic emission).}
\end{table*}

\clearpage

\begin{table}[tbh]
\begin{tabular}{l l l}
\hline 
Parameter/PWN Region & Compact PWN & Southern `Jet'  \\
\hline
$\Gamma$ & 1.1$^{+0.9}_{-0.7}$  & 1.4$^{+0.8}_{-0.9}$\\
Norm.\tablenotemark{a} & (2.2$^{+2.8}_{-1.2}$)$\times$10$^{-6}$ & (3.5$^{+5.6}_{-1.9}$)$\times$10$^{-6}$ \\
$\chi_{\rm \nu}^2$ (dof) & 1.23 (17) & 1.16 (17)  \\
$f_{\rm unabs}$ (0.5--7 keV) ergs~cm$^{-2}$~s$^{-1}$ & (1.9$^{+2.7}_{-1.0}$)$\times$10$^{-14}$  & (2.5$^{+3.9}_{-1.4}$)$\times$10$^{-14}$\\
\hline
\end{tabular}
\caption{Power law spectral fits to the PWN. The compact PWN refers to the ~6$^{\prime\prime}$$\times$15$^{\prime\prime}$  structure elongated in the north--south direction (see Fig.~4).
The southern `jet' corresponds to the new faint and extended
jet-like structure ($\sim$6$^{\prime\prime}$$\times$20$^{\prime\prime}$) shown
in Fig.~5. $N_H$ is frozen to  1.8$\times$10$^{22}$ cm$^{-2}$. Errors are 2$\sigma$.}
\tablenotetext{a}{Normalization at 1 keV in units of photons~cm$^{-2}$~s$^{-1}$~keV$^{-1}$.}
\end{table}

\clearpage

\begin{table}
\begin{tabular}{l l l l l l}
\hline 
Pulsar (PSR) & $\dot{E}$ (ergs~s$^{-1}$) & $B$ (10$^{13}$ G) & PWN? & $\frac{L_x (psr+pwn)}{\dot{E}}$ & Reference\\ 
\hline
J1846$-$0258 & 8.0$\times$10$^{36}$ & 4.9 & Y (X) & 0.018--0.0325\tablenotemark{a} $d_{6}^2$ & Kumar \& Safi-Harb (2008)\\
J1119$-$6127 &  2.3$\times$10$^{36}$ & 4.1 & Y (X) & 5$\times$10$^{-4}$$d_{8.4}^2$ & this work\\
J1734$-$3333 & 5.6$\times$10$^{34}$ & 5.2 & -- & -- & Morris et al. (2002)\\
J1718$-$3718 & 1.5$\times$10$^{33}$ &  7.4 & -- & 0.013--4.0$d_{4.5}^2$ & Kaspi \& McLaughlin
(2005) \\
J1814$-$1744 & 4.7$\times$10$^{32}$ & 5.5 & -- &  $<$1340\tablenotemark{b} $d_{10}^2$ & Pivovaroff et al. (2000)\\
J1819$-$1458\tablenotemark{c} & 2.9$\times$10$^{32}$ & 5.0 & -- & $\sim$12$d_{3.6}^2$ & Reynolds et al. (2006)\\ 
J1847$-$0130 & 1.7$\times$10$^{32}$ & 9.4 &  -- & $<$18--47\tablenotemark{d} $d_8^2$ & McLaughlin et al. (2003)\\ \hline
B1800$-$21 & 2.2$\times$10$^{36}$ & 0.4 & Y(X) & 10$^{-4}$$d_4^2$ & Kargaltsev et al. (2007)\\
J1357$-$6429 & 3.1$\times$10$^{36}$ & 0.8 & Y(X) & 6.5$\times$10$^{-5}$$d_{2.5}^2$ & Zavlin et al. (2007b)\\
\hline
\end{tabular}
\caption{The high-magnetic field pulsars sorted in order of decreasing spin-down power.  We used the ATNF pulsar catalog in compiling the pulsar properties and only selected pulsars with $B$$\gtrsim$4$\times$10$^{13}$ G (the QED value) (B is defined in footnote\#4, but see footnote\#13). The last two pulsars, shown only for comparison with \j1119,
 are rotation-powered pulsars with a `typical'
magnetic field, but with $\dot{E}$ and
 X-ray properties almost identical to PSR~J1119$-$6127.  The 4th column refers to the presence (Y) or absence  of a PWN in X-rays (X). Except for PSR~J1846$-$0258, none of these pulsars has a radio PWN associated with it. 
The 5th column refers to the ratio of the non-thermal X-ray luminosity of the combined pulsar and associated PWN (when detected in X-rays) to the pulsar's spin down power, 
scaled to the distance given in units of $d_x$, 
where $x$ is the adopted distance to the pulsar  in units of kpc.}
\tablenotetext{a}{the lower (upper) range corresponds to the 2000 (2006) $Chandra$ data; the pulsar has brightened by a factor
of $\sim$6 in 2006.}
\tablenotetext{b}{only an upper limit on the X-ray luminosity was estimated using the spectral properties of 1E~2259+586 (an AXP) as a template.}
\tablenotetext{c}{pulsar classified as a rotating radio transient; the
X-ray luminosity of the pulsar is uncertain by an order of magnitude.}
\tablenotetext{d}{only an upper limit was estimated using AXPs 4U0142+61 or 1E 2259+586 as templates.}
\end{table}

\clearpage

\begin{figure}
\epsscale{0.75}
\centerline {\plotone{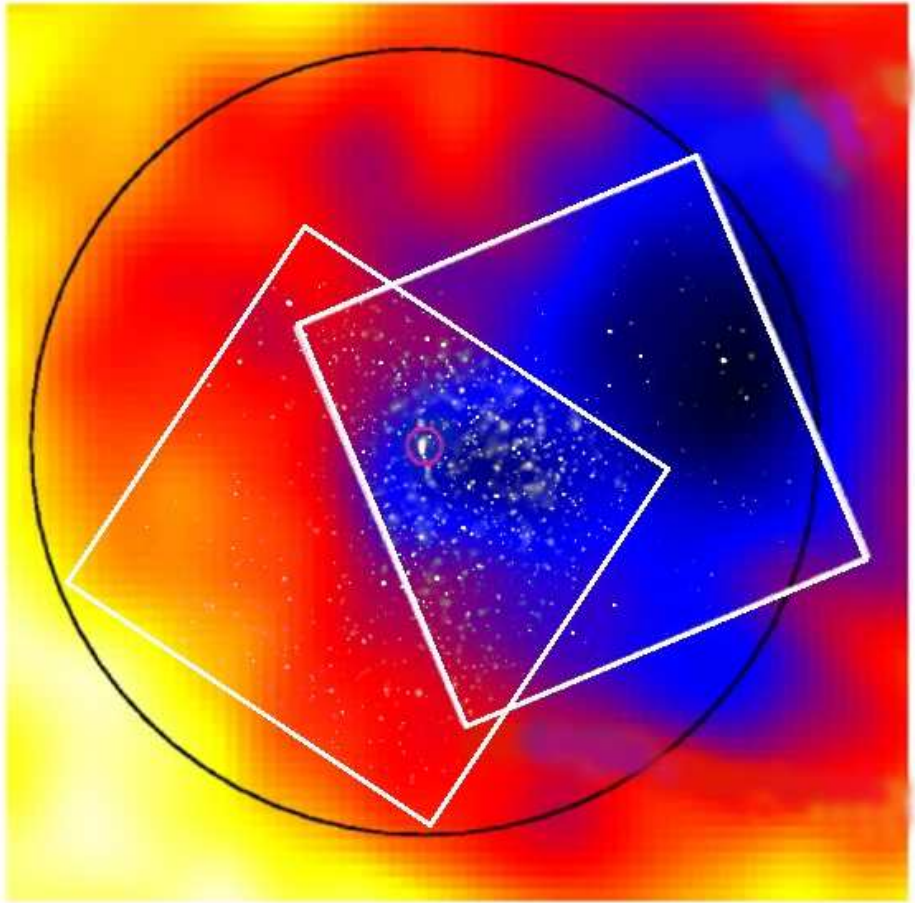}}
\caption{\label{1_01} 2.0--10.0 keV $\it{ASCA}$ image of SNR \g292,
smoothed with a Gaussian with $\sigma$=45$^{\prime\prime}$, with the
black circle showing the radio boundary (15$^{\prime}$ in diameter).
The superimposed squares mark the locations of 
the 8$^{\prime}$$\times$8$^{\prime}$ \chan's S3 chip during
AO-3 (right) and AO-5 (left), covering the western and eastern sides of
the remnant, respectively, with the grayscale image showing the
raw 0.5--10 keV emission seen with $Chandra$. The central 
circle marks the position of
PSR J1119$-$6127. North is up and east is to the left.}
\epsscale{1.0}
\end{figure}

\begin{figure}
\epsscale{0.75}
\centerline {\plotone{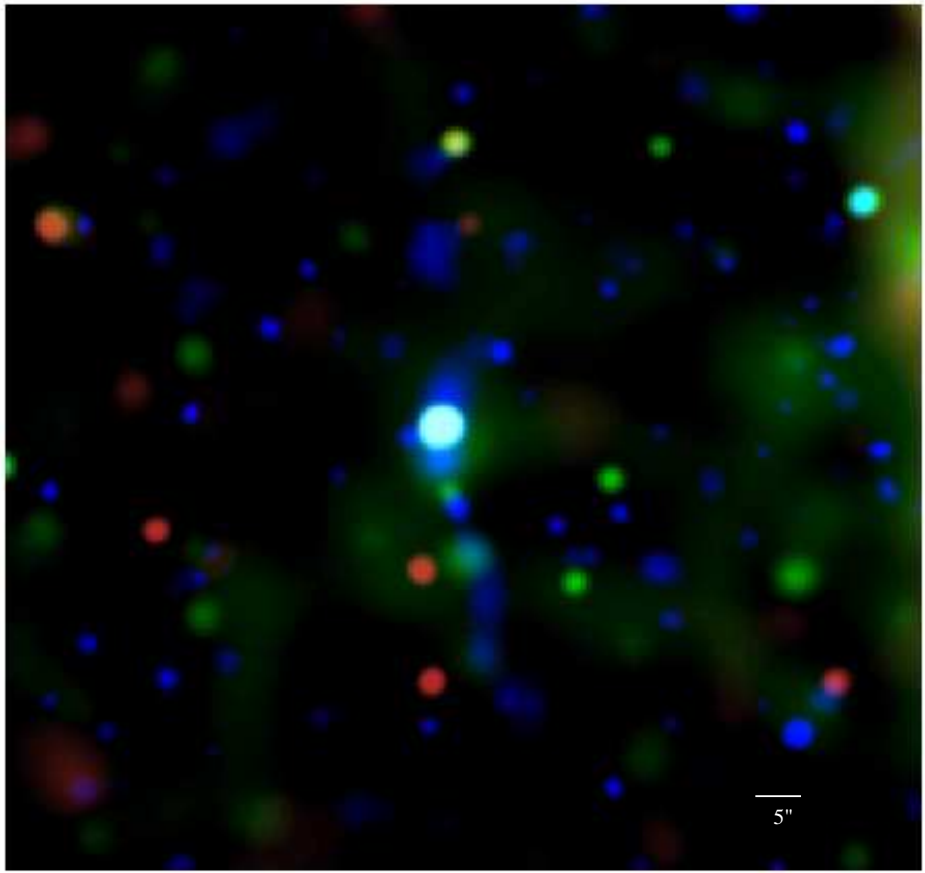}}
\caption{\label{1_02} The tri-color energy image of \j1119
and associated PWN. 
The image is 2.4$^{\prime}$$\times$2.3$^{\prime}$ in size
and is exposure corrected.
Red, green, and blue correspond to 0.5--1.15 keV, 1.15--2.3 keV, and
2.3--10.0 keV, respectively (see \S3.1 for details).}
\epsscale{1.0}
\end{figure}

\begin{figure}
\plottwo{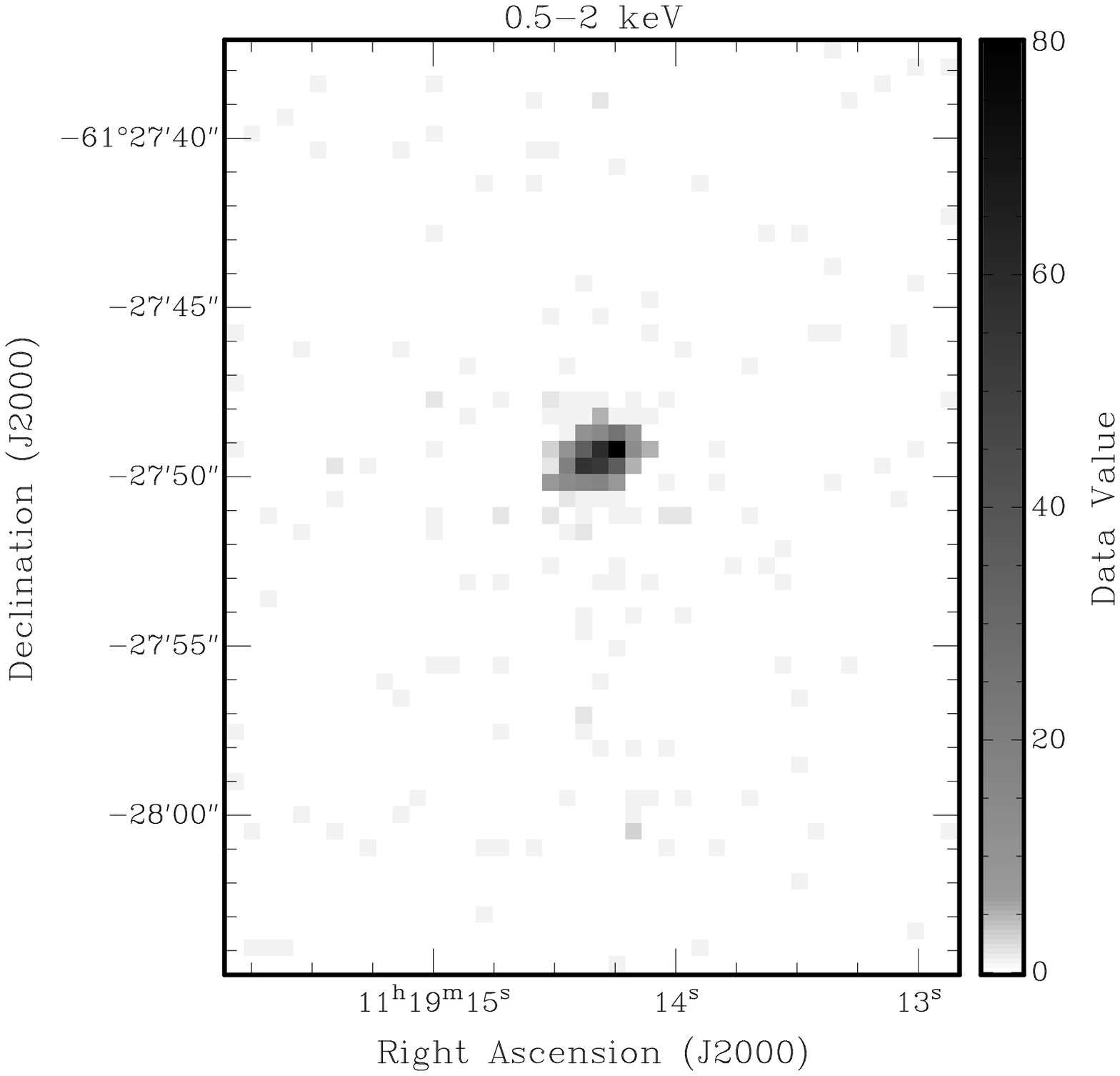}{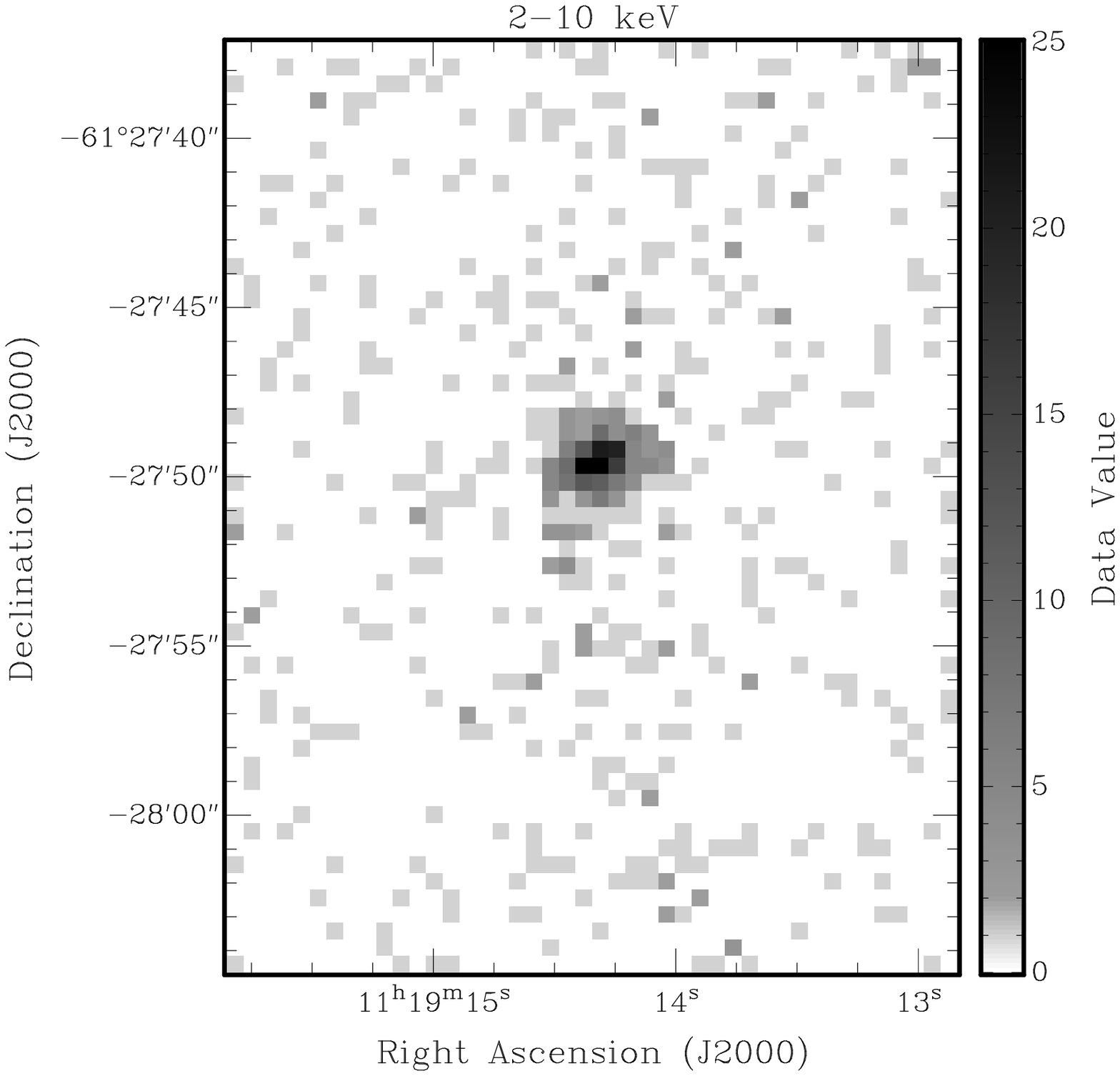}
\caption{The raw soft (0.5--2.0 keV) and hard (2.0--10 keV) images of PSR~J1119$-$6127
and associated nebula. These images were not background-subtracted.}
\end{figure}

\begin{figure}
\plottwo{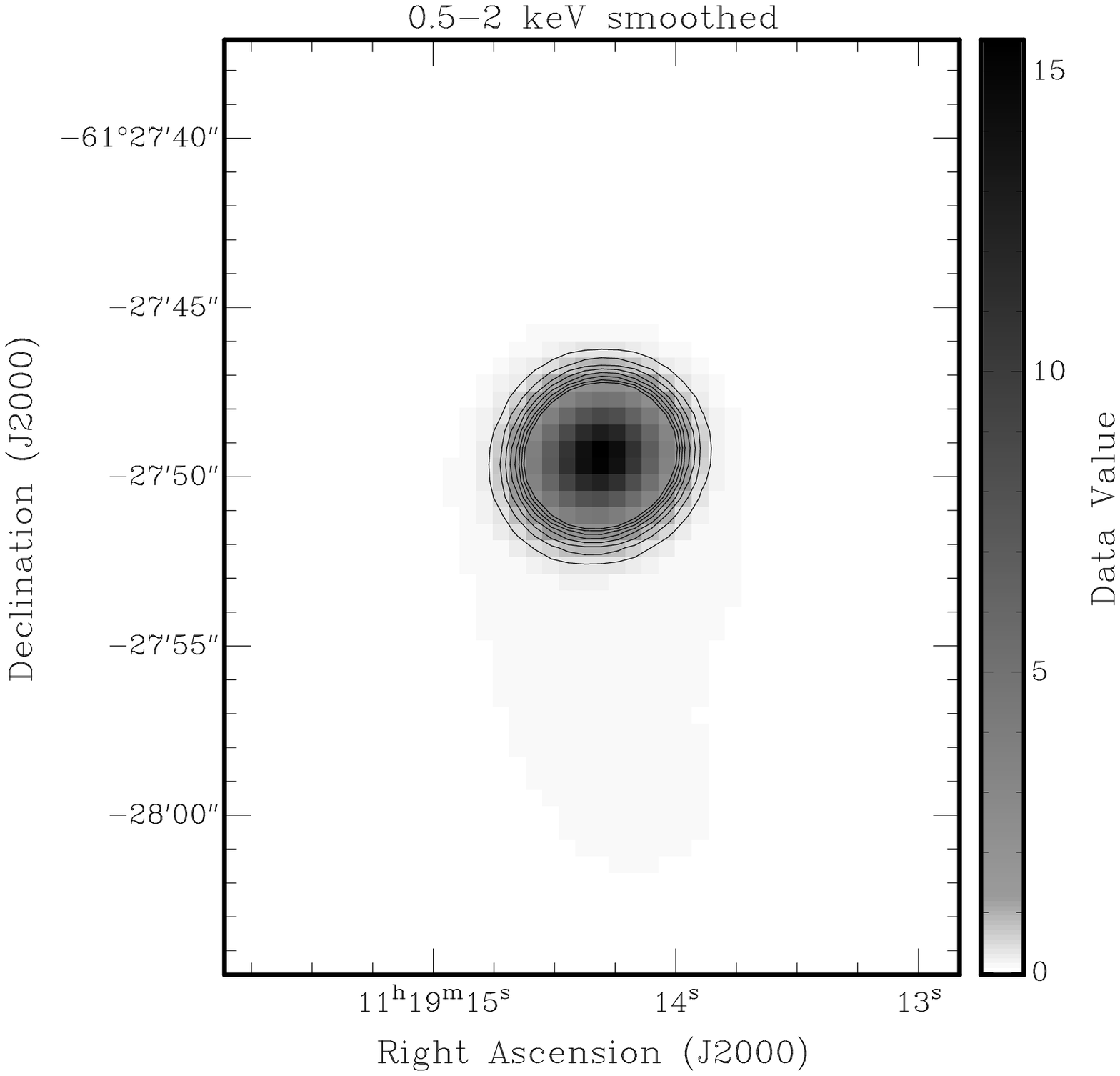}{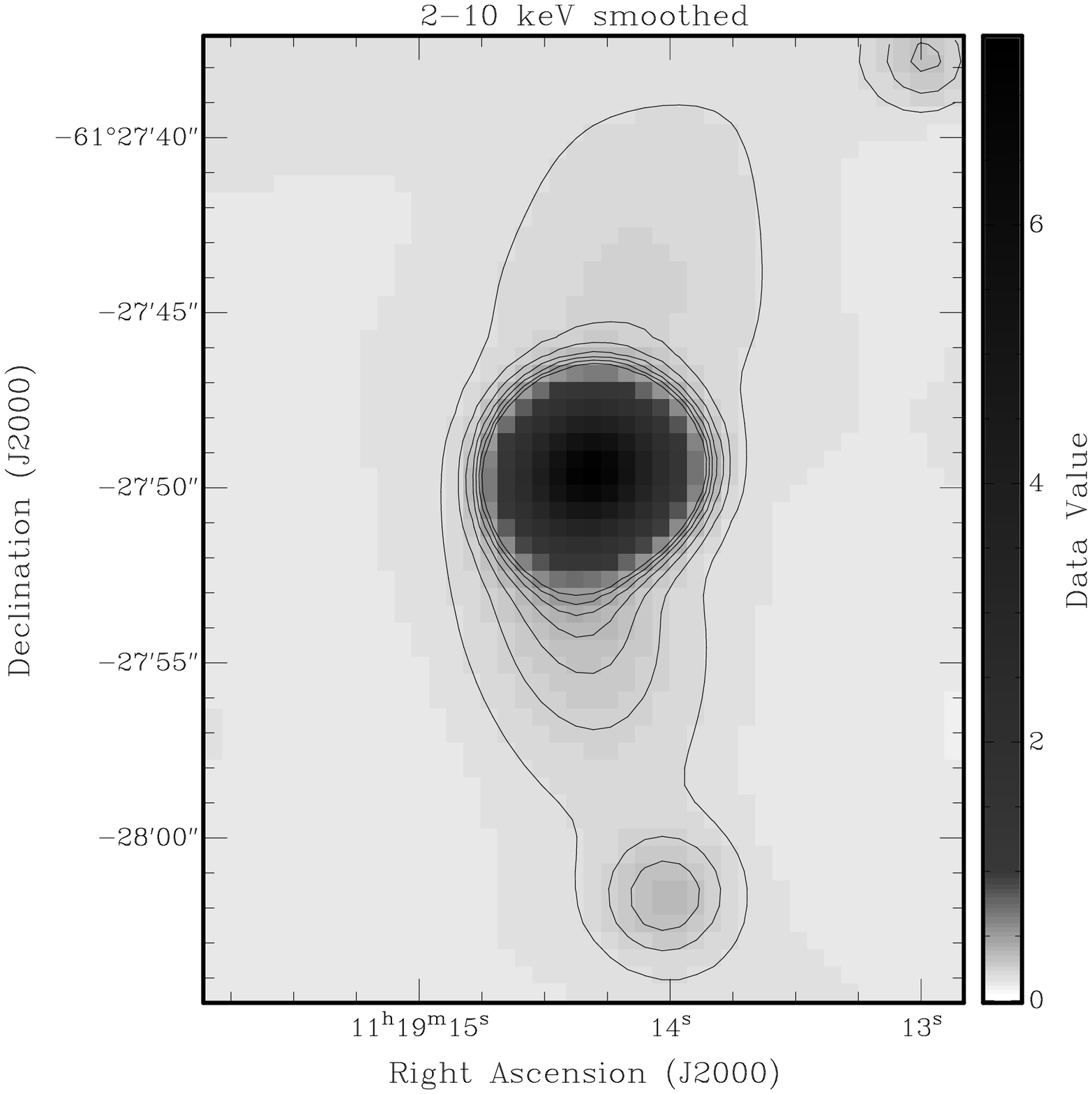}
\caption{ The same images as in Fig.~3, adaptively smoothed (with `csmooth') using a Gaussian with $\sigma$=1$^{\prime\prime}$ for
a significance of detection greater than 5 and up to $\sigma$=3$^{\prime\prime}$ for a
significance of detection down to 2. The PWN and its elongated morphology in the north-south direction is evident in the hard band image. Contours are overlayed on a linear scale, chosen to highlight the emission from the jets, and ranging from 2.5 to 17 counts per pixel in steps of 2 for the soft band image, and from 2.5 to 7 counts per pixel in steps of 0.75  for the hard band image. }
\end{figure}

\begin{figure}
\centerline{\plotone{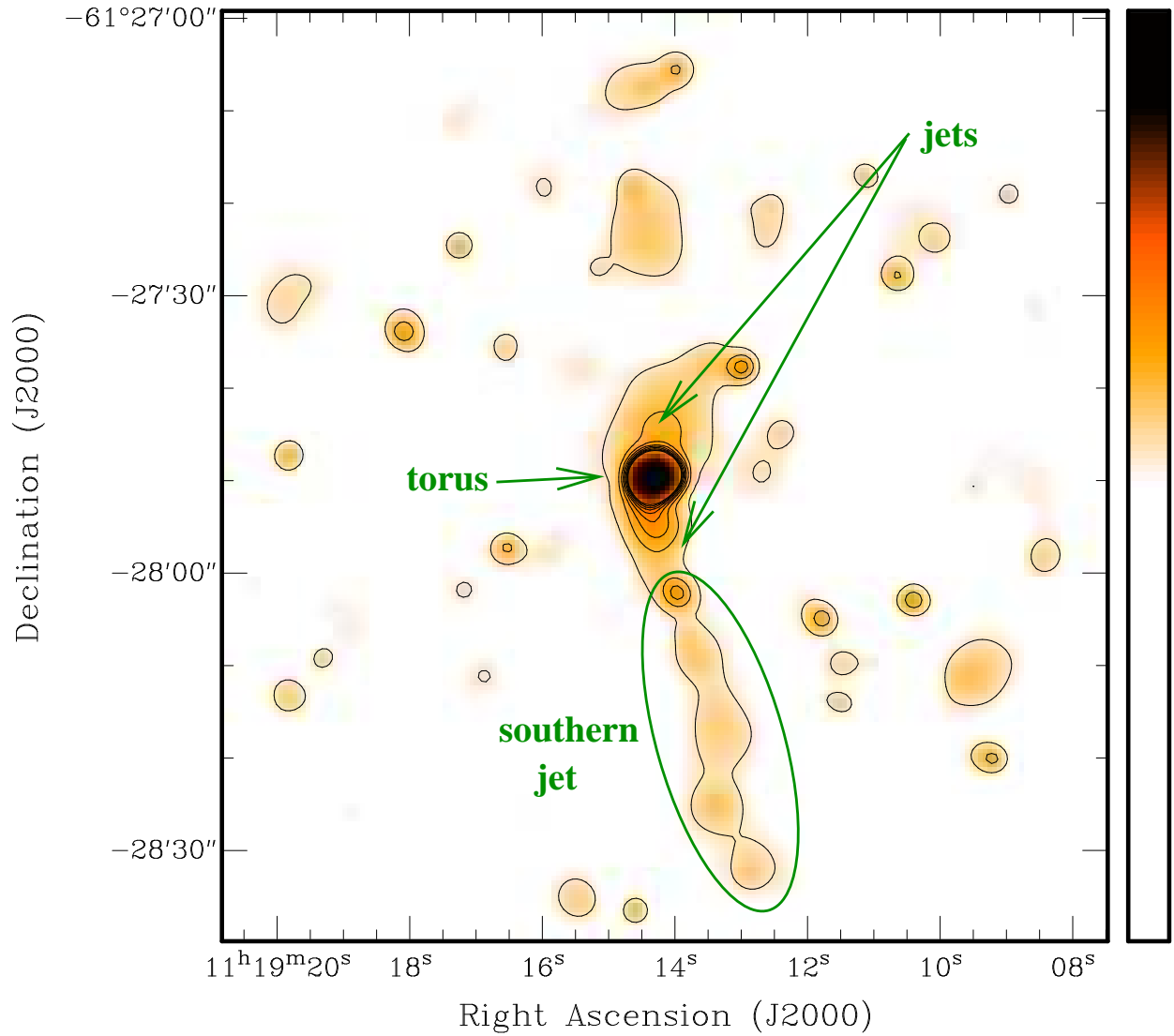}}
\caption{\label{1_03} The 2--10 keV  image of PSR J1119$-$6127 displaying the extended jet-like feature southwest of the pulsar. The image has been background-subtracted and smoothed as the previous figure.  The overlaid contours range from 1.4 to 10 counts per pixel in steps of 1 in linear scale and are
chosen to highlight the structure of the elongated jet.}
\end{figure}

\begin{figure}
\plotone{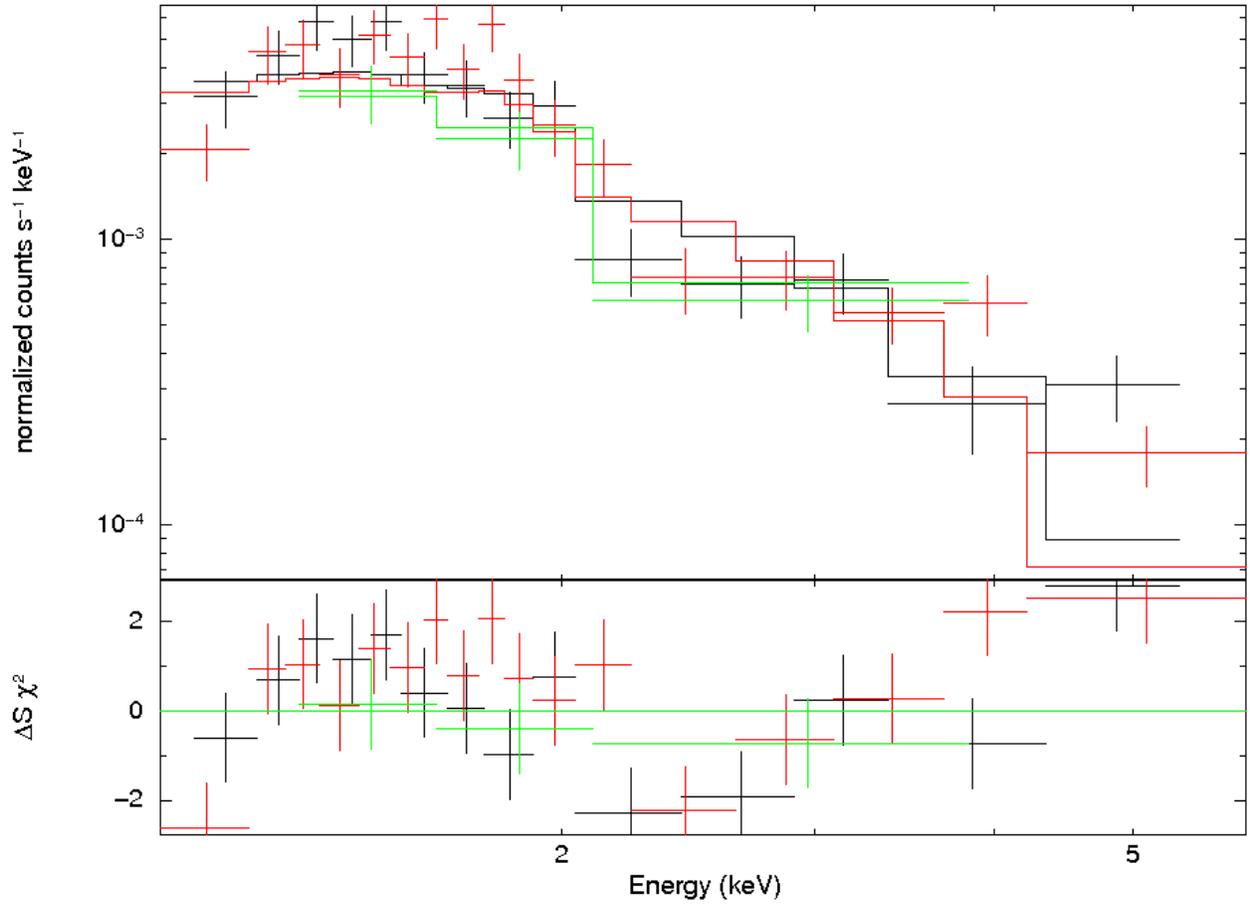}
\caption{The spectrum of PSR J1119$-$6127 fitted with the blackbody model,
illustrating the excess of emission above $\sim$3~keV unaccounted for by the blackbody model. 
The lower panel shows the ratio of data to plotted model. 
The 3 colors refer to the 3 observations used in our analysis ($obsid$ 2833 in black, 4676 in red, and 6153 in green).}
\end{figure}

\begin{figure}
\plotone{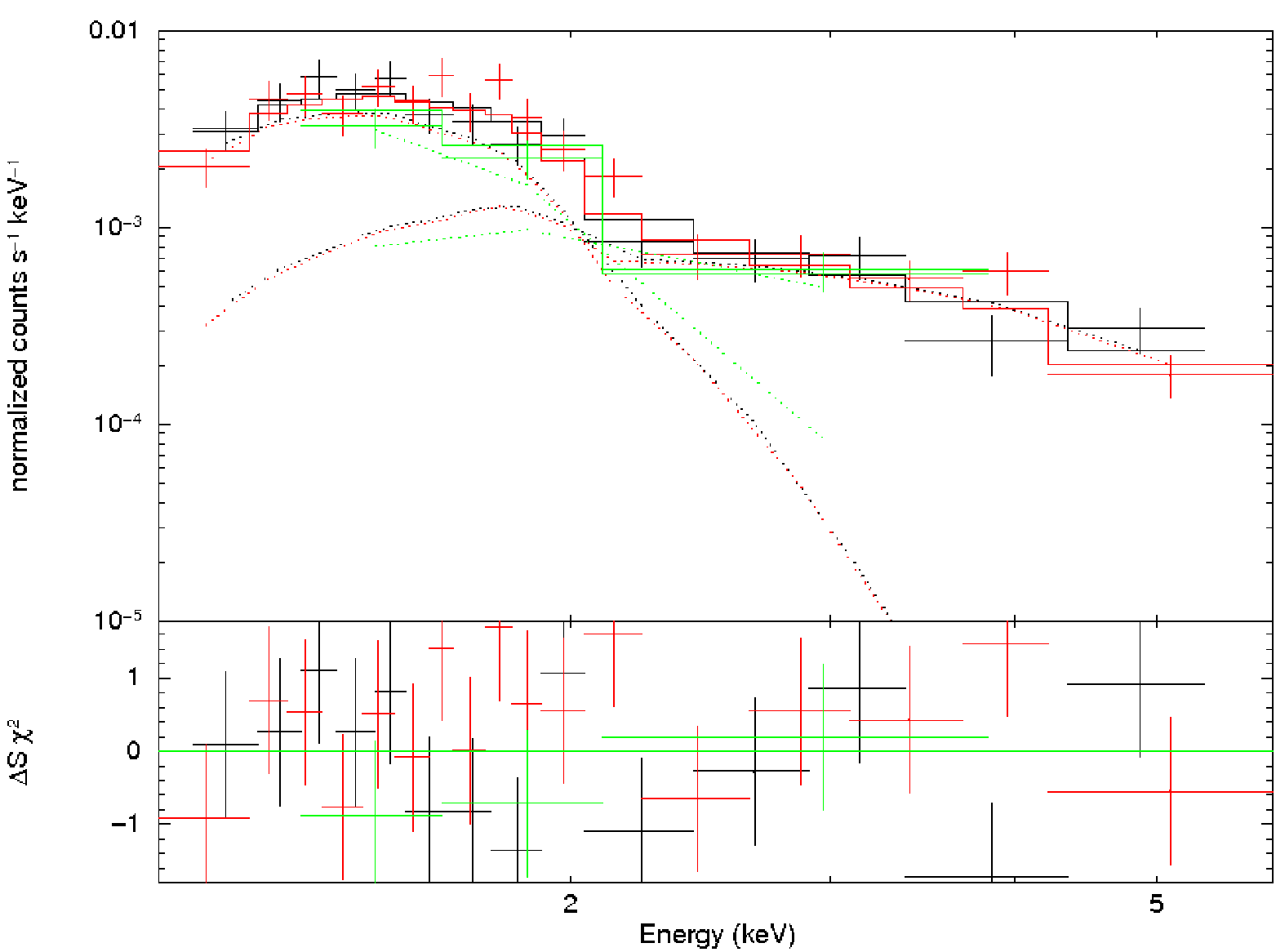}
\caption{The spectrum of PSR J1119$-$6127 fitted with the BB+PL model shown in Table~1. The lower panel shows the ratio of data to plotted model. The dotted lines show the contributions of the blackbody (soft) and power law (hard) components to the spectrum. The 3 colors refer to the 3
observations used in our analysis ($obsid$ 2833 in black, 4676 in red, and 6153 in green).}
\end{figure}


\begin{thebibliography}{}

\bibitem[]{501} Bucciantini, N. 2008, Advances in Space Research, 41, 491

\bibitem[]{} Camilo, F. 2008, Nature Physics, 4, 353

\bibitem[]{375} Camilo, F. et al.  2000, \apj, 541, 367


\bibitem[]{507} Camilo, F., Ransom, S. M., Halpern, J. P., Reynolds, J., Helfand, D. J., Zimmerman, N., \& Sarkissian, J. 2006,
Nature, 442, 892

\bibitem[]{510} Camilo, F.,  Ransom, S. M. , Halpern, J. P., \& Reynolds, J. 2007, \apj, 666, L93
 
\bibitem[]{512} Cordes, J. M., \& Lazio, T. J. W. 2002, preprint
(arXiv:0207156) 

\bibitem[]{376} Crawford, F. et al.  2001, \apj, 554, 152

\bibitem[]{516} Ebeling, H., White, D. A., Rangarajan, F. V. N. 2006, \mnras, 368, 65

\bibitem[]{518} Gaensler, B. M., \& Slane, P. O. 2006, \araa, 44, 17

\bibitem[]{520} Gaensler, B. M. et al.  2002, \apj, 569, 878

\bibitem[]{522} Gavriil, F. P., Gonzalez, M. E., Gotthelf, E. V., Kaspi, V. M., Livingstone, M. A., \& Woods, P. M. 2008, Science, 319, 1802
 
\bibitem[]{525} Gonzalez, M. E., \& Safi-Harb, S. 2003, \apj, 591, L143 (GSH03)

\bibitem[]{527} Gonzalez, M. E., \& Safi-Harb, S. 2005, \apj, 619, 856

\bibitem[]{529} Gonzalez, M. E., Kaspi, V. M., Camilo, F., Gaensler, B. M., Pivovaroff, M. J. 2005,  630, 489


\bibitem[]{534} Halpern, J. P., Gotthelf, E. V., Becker, R. H., Helfand, D. J., White, R. L. 2005,
\apj, 632, L29

\bibitem[]{537}	Kargaltsev, O. Y., Pavlov, G. G., Teter, M. A., \& Sanwal, D. 2003, New Astronomy
Reviews, 47, 487

\bibitem[]{540} Kargaltsev, O., Pavlov, G. G., \& Garmire, G. P. 2007, \apj, 660, 1413

\bibitem[]{542} Kaspi, V. M., \& McLaughlin, M. A. 2005, \apj, 618, L41

\bibitem[]{380_01} Kennel, C. F. \& Coroniti, F. V. 1984, \apj, 283, 694

\bibitem[]{380_02} Kumar, H. S., \& Safi-Harb, S. 2008, \apjl, 
678, L43

\bibitem[]{548} McLaughlin, M. A. et al. 2003, \apj, 591, L135

\bibitem[]{550} Mignani, R. P. et al. 2007, A\&A, 471, 265

\bibitem[]{552} Mirabel, I. F.,  \& Rodriguez, L. F. 1999, \araa, 37, 409

\bibitem[]{554} Morris, D. J. et al. 2002, MNRAS, 335, 275

\bibitem[]{556} Petre, R., Hwang, U., Holt, S. S., Safi-Harb, S., \& Williams, R. M. 2007, \apj, 662, 988

\bibitem[]{558} Pivovaroff, M. J., Kaspi, V. M., \& Camilo, F. 2000, \apj, 535, 379

\bibitem[]{386_01} Pivovaroff, M. J., Kaspi, V. M., Camilo, F., Gaensler, B. M., \& Crawford, F. 2001, \apj, 554, 161


\bibitem[]{} Reynolds, S. P. et al. 2006, 639, L71

\bibitem[]{562} Safi-Harb, S. 2008, in: 40 Years of Pulsars, Millisecond Pulsars, Magnetars and More. 
AIP Conference Proceedings, Volume 983, pp. 213-215

\bibitem[]{386_02} Slane, P. O., Helfand, D. J., \& Murray, S. S. 2002, \apj, 571, 45


\bibitem[] {} Woods, P. \& Thompson, C. 2006 in Compact stellar X-ray sources. Eds W. Lewin \& M. van der Klis. 
Cambridge Astrophysics Series, No. 39. Cambridge, UK: Cambridge University Press, p. 547 - 586

\bibitem[]{572} Zavlin, V. E., Pavlov, G. G., \& Shibanov, Y. A. 1996,  A\&A, 315, 141

\bibitem[]{574} Zavlin, V. E. 2007a, preprint (arXiv:0702426)

\bibitem[]{576} Zavlin, V. E. 2007b, \apj, 665, L143

\end{thebibliography}
\end{document}